\documentclass[acmtog]{acmart}
\acmSubmissionID{238}

\usepackage{booktabs} 
\usepackage{wrapfig}
\usepackage[ruled]{algorithm2e} 

\SetAlFnt{\small}
\SetAlCapFnt{\small}
\SetAlCapNameFnt{\small}
\SetAlCapHSkip{0pt}

\usepackage{xspace}
\makeatletter
\DeclareRobustCommand\onedot{\futurelet\@let@token\@onedot}
\def\@onedot{\ifx\@let@token.\else.\null\fi\xspace}
\def\eg{\emph{e.g}\onedot} 
\def\ie{\emph{i.e}\onedot}

\def\wrt{w.r.t\onedot} 
\def\etal{\emph{et al}\onedot}
\makeatother

\newcommand{\bB}{\mathbf{B}}

\newcommand{\bE}{\mathbf{E}}
\newcommand{\bF}{\mathbf{F}}

\newcommand{\bS}{\mathbf{S}}
\newcommand{\bT}{\mathbf{T}}

\newcommand{\bX}{\mathbf{X}}

\newcommand{\bb}{\mathbf{b}}

\newcommand{\bd}{\mathbf{d}}

\newcommand{\bff}{\mathbf{f}}
\newcommand{\bg}{\mathbf{g}}

\newcommand{\bn}{\mathbf{n}}

\newcommand{\bt}{\mathbf{t}}

\newcommand{\bx}{\mathbf{x}}

\newcommand{\bq}{\mathbf{q}}

\newcommand{\bZero}{\mathbf{0}}
\newcommand{\Lag}{\mathcal{L}}
\newcommand{\tran}{\mathsf{T}}
\newcommand{\dd}{\mathrm{d}}

\newcommand{\new}[1]
{
\textcolor{black}{#1}
}

\newcommand{\revision}[1]
{
\textcolor{black}{#1}
}

\begin{document}

\title{Neural Metamaterial Networks for Nonlinear Material Design}

\author{Yue Li}
\affiliation{%
  \institution{ETH Z{\"u}rich}
  \country{Switzerland}
}
\email{yue.li@inf.ethz.ch}

\author{Stelian Coros}
\affiliation{%
  \institution{ETH Z{\"u}rich}
  \country{Switzerland}
}
\email{stelian.coros@inf.ethz.ch}

\author{Bernhard Thomaszewski}
\affiliation{%
  \institution{ETH Z{\"u}rich}
  \country{Switzerland}
}
\email{bthomasz@ethz.ch}

\renewcommand\shortauthors{Li, Y. et al}

\begin{abstract}
Nonlinear metamaterials with tailored mechanical properties have applications in engineering, medicine, robotics, and beyond.
 While modeling their macromechanical behavior is challenging in itself, finding structure parameters that lead to ideal approximation of high-level performance goals is a challenging task.
In this work, we propose Neural Metamaterial Networks (NMN)---smooth neural representations that encode the nonlinear mechanics of entire metamaterial families. 
Given structure parameters as input, NMN return continuously differentiable strain energy density functions, thus guaranteeing conservative forces by construction.
Though trained on simulation data, NMN do not inherit the discontinuities resulting from topological changes in finite element meshes. They instead provide a smooth map from parameter to performance space that is fully differentiable and thus well-suited for gradient-based optimization. On this basis, we formulate inverse material design as a nonlinear programming problem that leverages neural networks for both objective functions and constraints.
We use this approach to automatically design materials with desired strain-stress curves, prescribed directional stiffness and Poisson ratio profiles. We furthermore conduct ablation studies on network nonlinearities and show the advantages of our approach compared to native-scale optimization.

\end{abstract}
%


\setcopyright{acmlicensed}
\acmJournal{TOG}
\acmYear{2023} \acmVolume{42} \acmNumber{6} \acmArticle{1} \acmMonth{12} \acmPrice{15.00}\acmDOI{10.1145/3618325}


\begin{teaserfigure}
    \centering
    \includegraphics[width=\linewidth]{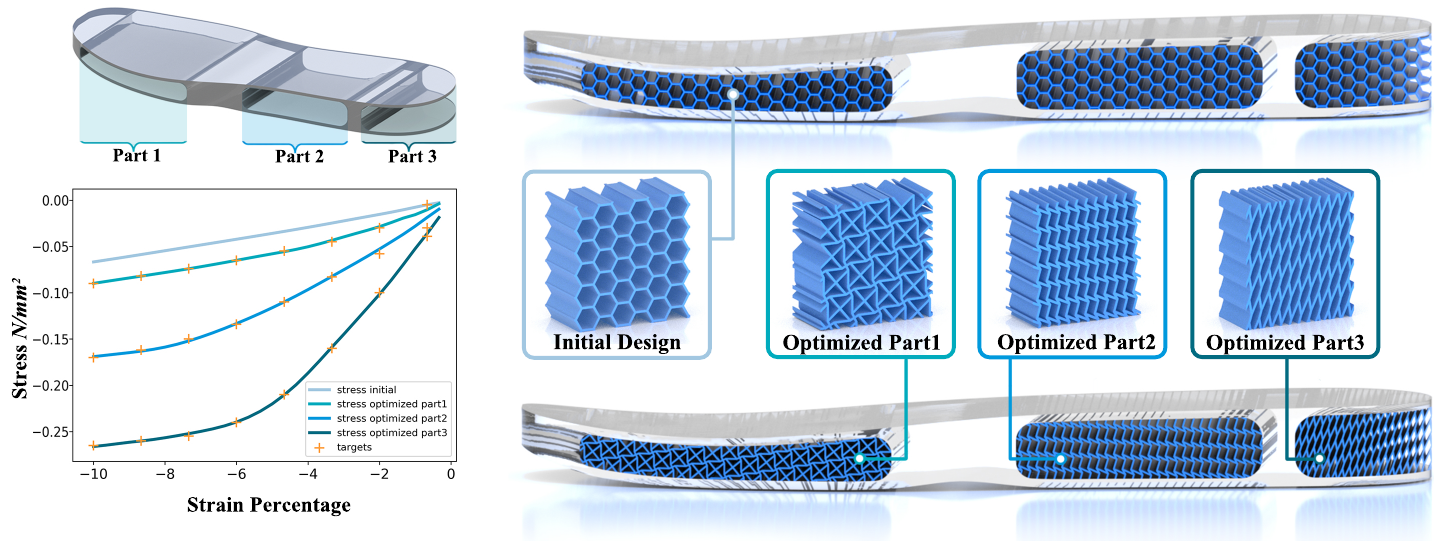}
    \caption{
    Inverse design with Neural Metamaterial Networks. For this custom shoe design, we aim to find metamaterials that best approximate given nonlinear stiffness targets in the forefoot, midfoot, and heel regions as shown on the \textit{left}. Starting from a homogeneous hexagonal pattern, our method leverages a differentiable neural representation to optimize over entire metamaterial families, resulting in structures that closely track the desired strain-stress curves.
    }
    \label{fig:teasor}
\end{teaserfigure}
\maketitle
\section{Introduction}
%
%
Thanks to their programmable microstructures, mechanical metamaterials boast a wide range of macroscopic properties.
There has recently been growing interest in flexible metamaterials that offer tailored mechanical properties in the finite-deformation regime \cite{bertoldi2017flexible}. Their ability to sustain large deformations makes these metamaterials interesting and useful for myriad applications including micro-electromechanical systems and precision instruments, functional sportswear and rehabilitation medicine, as well as robotics and architecture.
A common trait shared by these examples is that they operate at very large deformations, which means that nonlinear effects in both geometry and material become predominant: beams in the microstructure rotate and straighten under stretching, whereas compression leads to buckling and self-contact. 
\par
To design in this nonlinear setting, we must be able to determine structure parameters that lead to an ideal approximation of given high-level performance goals. This problem is challenging for two main reasons. First, creating a nonlinear macromechanical model for a given metamaterial is a highly nontrivial task. While Finite Element simulations can be used to generate large amounts of stress-strain data, combining these data into an accurate representation that preserves first principles such as objectivity, convexity, and integrability is difficult unless limiting assumptions on material symmetries and nonlinearities are introduced. 
Second, changing a material's structure parameters will entail changes in the simulation meshes. While these changes are often smooth, there will inevitably be singular points at which mesh topology must change to accommodate changes in geometry. 
These topology changes, and the corresponding discontinuities in simulation derivatives, are challenging the robustness of gradient-based optimization methods.
\par
In this work, we propose Neural Metamaterial Networks (NMN) to model the nonlinear macromechanical behavior of entire metamaterial families. Given structure parameters as input, NMN return a continuously differentiable strain energy density function, which guarantees conservative forces by construction.
Though trained on simulation data, NMN do not inherit the discontinuities resulting from topological changes in finite element meshes. They instead provide a smooth map from parameter to performance space that is fully differentiable and thus ideally suited for gradient-based optimization. Using neural metamaterial networks as a basis, we formulate inverse material design as a nonlinear programming problem that leverages neural networks for both the objective function and the constraints.
\par
We demonstrate our approach on a set of inverse material design examples where we automatically compute structure parameters that best approximate desired strain-stress curves, prescribed directional stiffness, and Poisson ratio profiles. We furthermore conduct ablation studies on our choice of network nonlinearities, and show the advantage of computing gradient information from our neural representation compared to simple finite-differencing on native scale simulations.

\section{Related Work}
\begin{figure*}[t!]
    \centering
    \includegraphics[width=\linewidth]{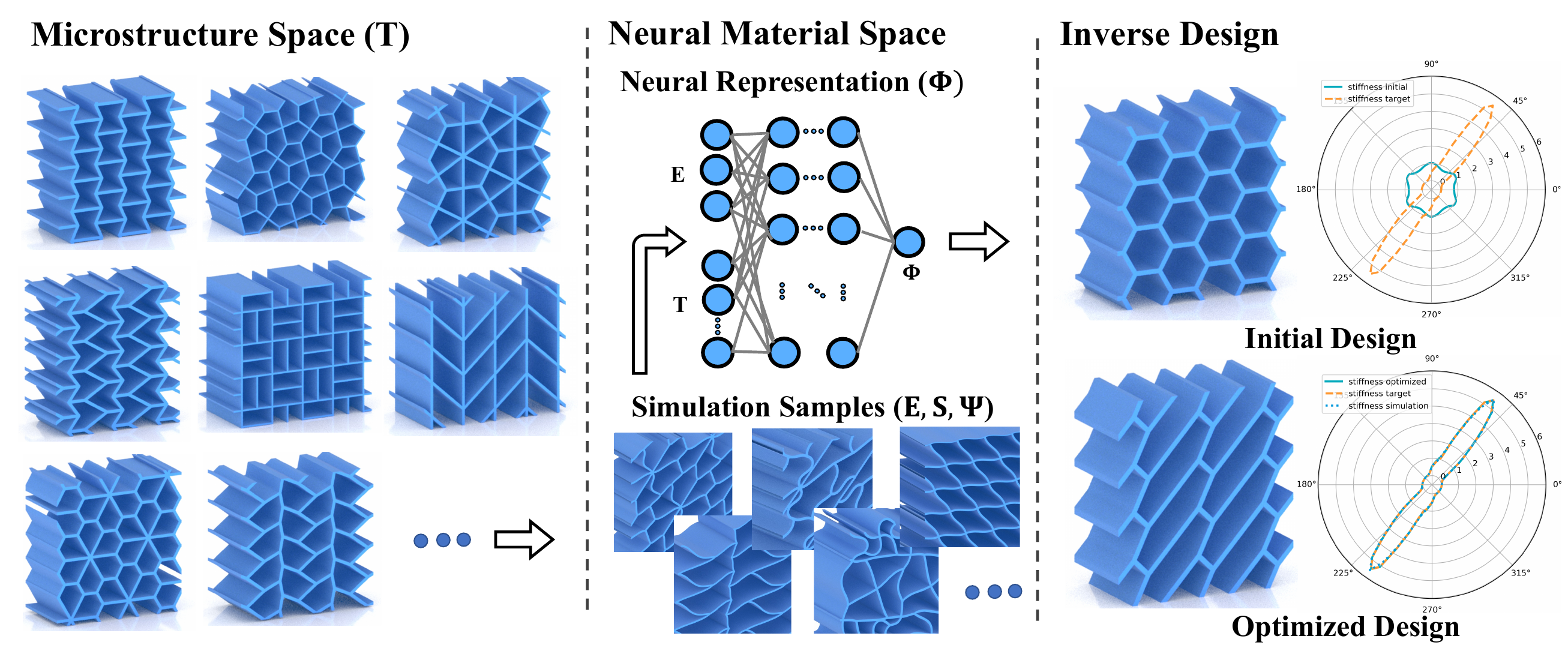}
    \caption{Pipeline. Our metamaterials leverage families of isohedral tiling patterns as geometries. 
    For a given set of tiling parameters $\bT$, we sample the nonlinear response of the corresponding unit cell by computing equilibrium configurations for uni- and bi-axial states of strain with different directions and magnitudes. 
    Using these simulation data, we perform homogenization on the unit patch to extract macroscale descriptions, \ie Green strain $\bE$, Second Piola-Kirchhoff stress $\bS$, and the energy density $\Psi$. 
    To capture the constitutive relationship for an entire metamaterial family, we train a deep neural network to learn the map from strain $\bE$ and tiling parameters $\bT$ to energy density functions $\Phi$. 
    The resulting Neural Metamaterial Networks are ideally suited for inverse material design with analytical derivatives. 
}
    \label{fig:pipeline}
\end{figure*}
\paragraph{Structured Metamaterials.}
Through careful tuning of their micro-scale geometry, metamaterials can cover a wide range of marcomechanical properties.
Designing metamaterials with desired macromechanical properties is a problem that spans many areas explored in graphics, including
geometry processing, physics-based modeling, and computational fabrication. Consequently, there has recently been a surge in works that explore various types of metamaterials \cite{schumacher2018mechanical,martinez2019star,Leimer2020,leimer2022analysis,li2022programmable,martinez2017orthotropic,tricard2020freely,efremov20213d}.
For example, Schumacher \etal~\shortcite{schumacher2018mechanical} characterize the in- and out-of-plane behavior of structured sheet materials. 
Li \etal~\shortcite{li2022programmable} proposed a methodology for creating 3D-printable, weave-like materials with desired mechanical properties.
Based on star-shape metrics, Mart\'{i}nez \etal~\shortcite{martinez2019star} and Efremov \etal~\shortcite{efremov20213d} introduce spatially graded meta-materials that cover a wide range of elastic properties.
Mart\'{i}nez \etal~\shortcite{martinez2017orthotropic} leverage procedural Voronoi foams to efficiently create orthotropic microstructures.
Moreover, Thibault \etal~\shortcite{tricard2020freely} generate freely-orientable microstructures with locally controllable anisotropy.
While the above methods explore specific types of metamaterials, we address the fundamental problem of generating smooth representations of nonlinear macromechanical properties for entire metamaterial families---a problem that is at the core of inverse metamaterial design.

\paragraph{Nonlinear Material Design.}

Vast parameter spaces for complex, nonlinear materials require inverse design approaches for efficient, performance-driven navigation.
With the goal of matching desired deformation behaviors, Bickel \etal~\shortcite{bickel2010design} optimize for the best arrangements of layered microstructures.
By varying beam density and radii, Mart\'{i}nez \etal~\shortcite{martinez2016procedural} find 3D Voronoi foams for prescribed Young's modulus targets.
Zehnder \etal~\shortcite{zehnder2017metasilicone} optimize for the sizes, numbers, and locations of stiff inclusions to generate silicone metamaterials with desired mechanical properties.
Panetta \etal~\shortcite{panetta2015elastic} introduce a library of elastic truss-like structures and a two-stage shape optimization process to find the best matching structure for a given target elasticity tensor. 
Building on this approach, Panetta \etal~\shortcite{panetta2017worst} further perform shape optimization to reduce peak stresses for worst-case loads. 
Schumacher \etal~\shortcite{schumacher2015microstructures} find a set of microstructures with smoothly varying material properties,
and Tozoni \etal~\shortcite{tozoni2020low} map material parameters to families of rhombic microstructures.
Our method shares the same high-level goal of finding metamaterials that best approximate macromechanical performance goals. However, instead of interpolating strain-stress samples in parameter space, we leverage deep neural networks to learn energy density functions for a large range of deformations across entire metamaterial families.
\par
Previous work has successfully used sampling-based strategies for inverse metamaterial design~\cite{panetta2015elastic,schumacher2015microstructures,chen2018computational} in the context of linear elasticity. Whereas linear materials are described by a single elasticity tensor, our work targets nonlinear materials, which require significantly more data (\ie, multiple stress-strain samples). While interpolation between linear elasticity tensors can lead to reasonable behavior, it is unclear how to interpolate between nonlinear materials.
Close to our work that also uses isohedral tilings, Schumacher \etal~\shortcite{schumacher2018mechanical}, constructs stiffness profiles in a pre-processing step and retrieves the closest matches for given targets at runtime. Whereas they only support isotropic strains with two pre-defined magnitudes, our method extends to arbitrary states of strain with a continuous range of magnitudes. More importantly, Schumacher \etal only provide a few snapshots of a material’s performance whereas our method offers a complete and smooth material model that can be used for inverse design and forward simulation.
\par
An alternative strategy for metamaterial design is to use topology optimization~\cite{wang2014design,zhu2017two,andreassen2014design,clausen2015topology,behrou2021topology,nakshatrala2013nonlinear,allaire2004structural,wang2014topological,chen2018computational}. This approach can automatically discover optimal microstructures for given performance goals. However, whereas topology optimization returns a single microstructure for a given target behavior, our approach explores entire metamaterial families. 
\revision{While another stream of research 
explores triply periodic minimal surfaces as a design space for microstructures for two scale topology optimization~\cite{feng2022stiffness,zhang2022tpms,xu2023topology}, they target linear elasticity and small displacements, whereas we address nonlinear elasticity for large deformations. Moreover, we use machine learning for macromechanical characterization.}


\paragraph{Differentiable Simulation}
\new{A recent stream of graphics research focuses on differentiable simulation, which requires derivatives of simulation outputs \wrt input parameters. Whereas some works rely on auto-differentiation~\cite{hu2019chainqueen,hu2019difftaichi,jatavallabhula2021gradsim}, others employ the adjoint method to compute the analytical derivative~\cite{geilinger2020add,panetta2019x,tozoni2021optimizing,liang2019differentiable,panetta2015elastic,panetta2017worst,chen2021bistable}. Panetta \etal~\shortcite{panetta2015elastic,panetta2017worst} optimize for structure parameters using analytical shape derivatives obtained from a level set formulation. Different from their work, which requires re-meshing and native-scale simulation per optimization step, we propose a novel learning-based approach that removes the need for native-scale simulation and re-meshing during inverse design. Similar in spirit to previous work that replaced simulation with a learned model for optimization~\cite{tymms2020appearance,wang2022differentiable}, we leverage deep neural networks as analytically differentiable surrogates for nonlinear metamaterial design. }

\paragraph{Data-driven Constitutive Model.}

Modeling material behavior is a problem that has received much attention from the graphics community in recent years. One stream of research focuses on fitting material parameters of existing models to real-world data~\cite{becker2007robust,bickel2009capture,destrade2017methodical,hahn2019real2sim}. Another line of works explores dedicated material models, \eg for cloth~\cite{wang2011data,miguel2012data} and hyperelastic solids~\cite{Martin11EBEM,li2015stable,xu2015nonlinear,miguel2016modeling}. 
Most similar to our approach are methods based on neural networks~\cite{Wang20Learning,Masi21TANN,le2015computational,linka2021constitutive,li2022plasticitynet}. 
While machine learning approaches for modeling constitutive relationships have been explored since the early 90s~\cite{ghaboussi1991knowledge,jung2006neural}, the vast majority do not guarantee conservative forces. For example, Wang \etal~\shortcite{Wang20Learning} learn corrective stresses for nominal models from simulation data. 
However, as stress corrections do not derive from a potential, they are generally not conservative. %
A notable exception is the work by Masi \etal~\shortcite{Masi21TANN}, who propose a method for learning constitutive models using neural networks with guaranteed thermodynamical consistency. 
Whereas Le \etal~\shortcite{le2015computational} directly map linear strain tensors to energy density, Linka \etal~\shortcite{linka2021constitutive} precompute generalized strain invariants and then learn an energy density function whose derivatives best approximates nonlinear stress-strain data.  
Beyond the purely elastic regime, Li \etal~\shortcite{li2022plasticitynet} learn energy representations for metal, sand, and snow plasticity. 
Similar to these works, our method also leverages deep neural networks to learn constitutive models from simulation data. To the best of our knowledge, however, our method is the first to model entire families of metamaterials using neural networks.
%

\paragraph{Homogenization.}
Computational homogenization~\cite{bensoussan2011asymptotic,guedes1990preprocessing,geers2010multi} aims to extract the macromechanical behavior from micro-scale geometry.
Within the graphics community, homogenization theory has been explored for distilling coarse-scale material properties from fine-scale simulations~\cite{kharevych2009numerical,nesme2009preserving,chen2015data} as a means to reduce computation times.
\par
Using different computational models for micro- and macro-scale simulations, Sperl \etal~\shortcite{sperl2020homogenized} conduct homogenization on yarn-level simulation data to enable efficient knitwear simulation using a macro-scale shell solver.
Most closely related to our work are methods that use homogenization to compute the macromechanical properties of metamaterials ~\cite{schumacher2015microstructures,panetta2015elastic,schumacher2018mechanical,li2022programmable}.
While we also compute homogenized properties from simulation data distributed across strain space, we learn a continuous, analytically-differentiable energy density function whose derivatives best approximate the simulated strain-stress behavior.


\section{Method}
In this section, we describe the machinery required for nonlinear metamaterial design using Neural Metamaterial Networks. See also Fig.~\ref{fig:pipeline} for an overview.
We use finite element simulation to model the nonlinear mechanics of metamaterials at their native scale (Sec.~\ref{sec:method_sim}), and extract corresponding macromechanical properties using concepts from homogenization theory (Sec.~\ref{sec:method_homo}).
We use this data to train Neural Metamaterial Networks, \ie, deep neural networks that smoothly map structure parameters to macromechanical descriptions in the form of energy density functions and corresponding nonlinear stress-strain relationships (Sec.~\ref{sec:method_ncm}).
Finally, we demonstrate the potential of Neural Metamaterial Networks for a set of inverse design tasks, leveraging the analytical derivatives of the neural networks (Sec.~\ref{sec:method_inv_design}).
%
\subsection{Isohedral Tilings}
\begin{figure}[h!]
    \centering
    \includegraphics[width=\linewidth]{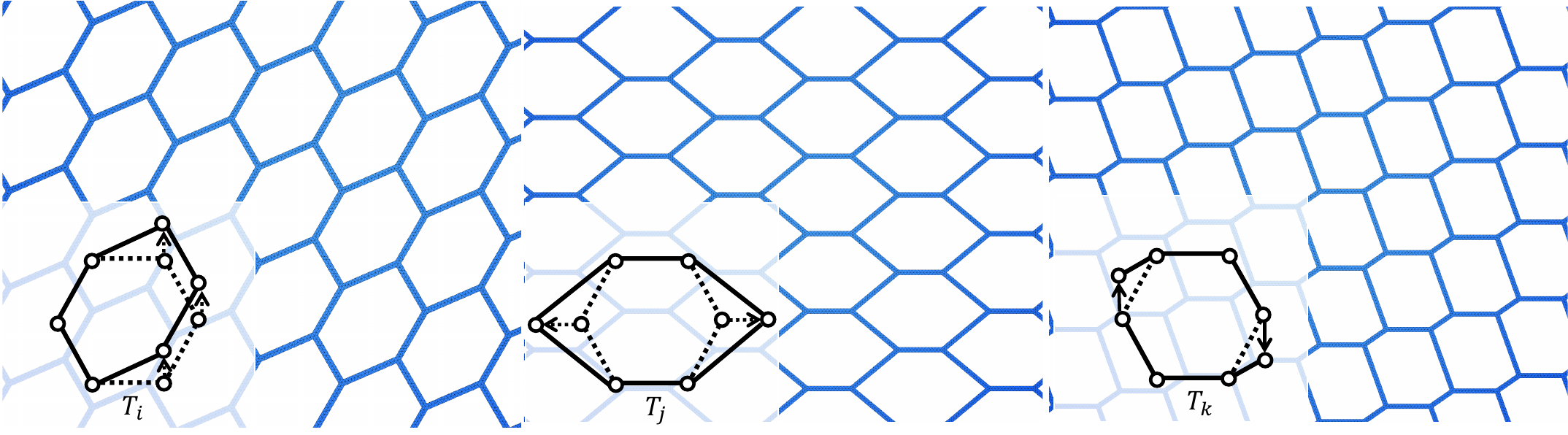}
    \caption{Tiling parameters of family IH01. Varying structure parameters leads to continuous changes in the basic tiling geometry (black hexagon). The effect of each parameter in this family is shown in the insets.}
    \label{fig:tiling_params}
\end{figure}

Our method targets parametric metamaterial families that are described through a set of variables defining the material's microstructure. Here we focus on isohedral tilings, a particular class of parametric families that span a large and diverse space of two-dimensional patterns \cite{kaplan2000escherization,schumacher2018mechanical}\new{\footnote{See also https://isohedral.ca/software/tactile/}.} An isohedral tiling family consists of patterns made by repeating a single base shape without gaps and overlaps. The geometry of the base shape is controlled by a number of continuous structure parameters $\bT$ (see Fig.~\ref{fig:tiling_params}). 
There are no topology changes within families, but topologies generally vary across families. We extrude these patterns out of their plane to obtain three-dimensional materials that are simple to manufacture using, \eg, 3D printers based on fused filament fabrication. 

\subsection{Microscale Simulation}
\label{sec:method_sim}
Since we are only interested in their in-plane behavior, we assume that loads along the normal direction are negligible. The corresponding plane stress assumption allows for an entirely two-dimensional treatment. 
For a given set of tiling parameters $\bT$, we first assign a width to all the tiling edges and then discretize them with quadratic triangle elements for simulation. 
We perform simulation on a periodic patch made of a $2\times2$ tiling of the minimal unit cell, see also Fig.~\ref{fig:periodic_unit}. 
\begin{figure}[h!]
    \centering
    \includegraphics[width=\linewidth]{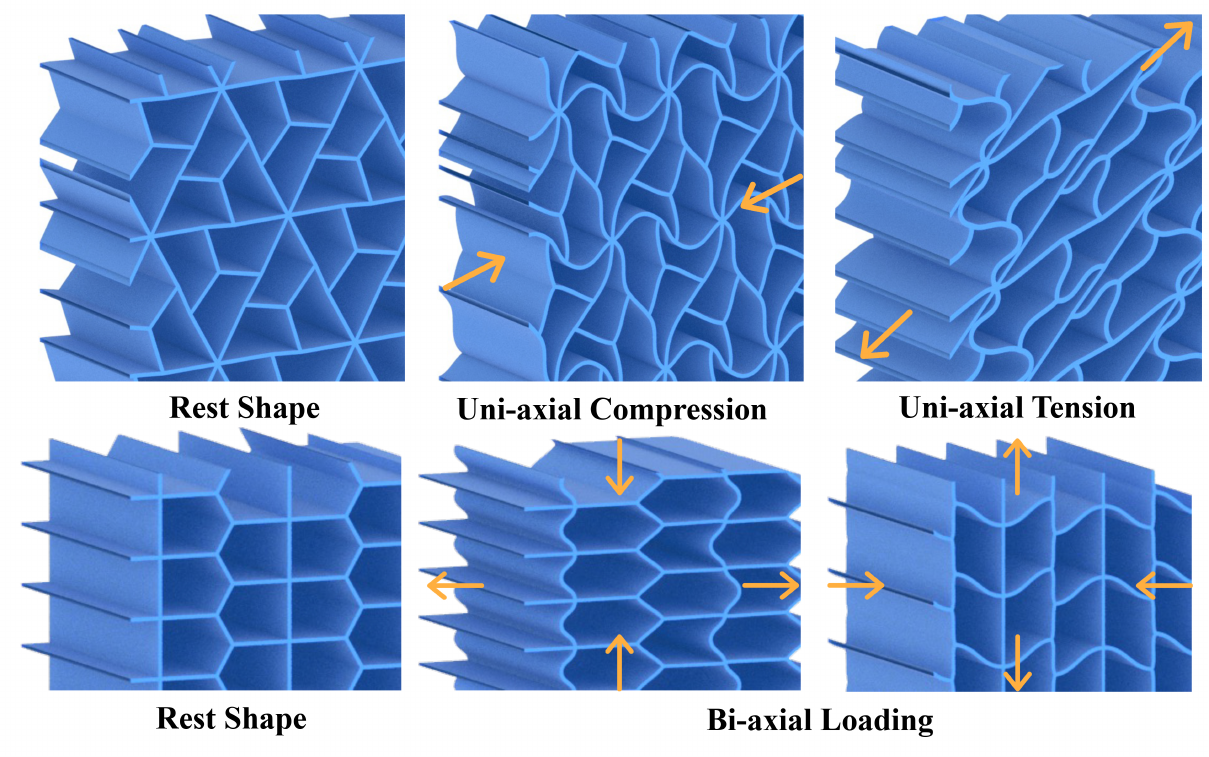}
    \caption{Simulation Samples. Here we show the static equilibrium configurations for uni- and bi-axial loading as indicated in orange. As can be seen from these images, our simulation framework reliably captures the substantial nonlinear buckling behaviors without any self-intersections.}
    \label{fig:simulation}
\end{figure}
\par
We compute static equilibrium configurations for given macroscopic strains by minimizing the potential
\begin{equation}
\begin{aligned}
    E_{\mathrm{total}} &= E_{\mathrm{elastic}}(\bx_\mathrm{r}(\bT), \bX_\mathrm{r}(\bT)) \\ &+ E_{\mathrm{strain}}(\bx_\mathrm{r}(\bT), \bX_\mathrm{r}(\bT)) + E_{\mathrm{contact}}(\bx_\mathrm{r}(\bT)) \ ,
\end{aligned}
\end{equation}
where $\bx_\mathrm{r}$ and $\bX_\mathrm{r}$ are the reduced simulation degrees of freedom (DoF) in the current and rest configurations, respectively. We define $\bx_r := (\bx, \bt_{ij}, \bt_{kl})$, where $\bt_{ij}$ and $\bt_{kl}$ are the translational DoF, with which the vertex positions on one side of the periodic patch is obtained from vertices on the other side. Therefore $\bx$ excludes vertices on two adjacent boundaries of the unit patch. With this treatment, periodicity is imposed implicitly as a hard constraint.
We use a standard incompressible Neo-Hookean material~\cite{bonet1997nonlinear} for the elastic potential $E_{\mathrm{elastic}}$, setting its Young's modulus and Poisson ratio according to the specifications of the printing filament.
We impose uni- and bi-axial strain conditions through penalty function $E_{\mathrm{strain}}$, asking that displacements of corresponding boundary vertices comply with target deformations.
Specifically, we define $E_{\mathrm{strain}}$ as
\begin{equation}
    E_{\mathrm{strain}} = \omega_\epsilon \sum_i (\bb_i^T \bd_\alpha - E_\alpha \bB_i^T\bd_\alpha)^2 + (\bb_i^T \bn_\alpha - E_\alpha^\perp \bB_i^T\bn_\alpha)^2\ ,
\end{equation}
where $\omega_\epsilon$ is the penalty weight, $\bd_\alpha$ is the direction vector
\begin{equation}
    \bd_\alpha = [\cos(\alpha), \sin(\alpha)]^{\tran} \ ,
\end{equation}
and $\bn_\alpha$ is the vector orthogonal to it. Furthermore, $E_\alpha$ and $E_\alpha^\perp$ are the corresponding strain magnitude in the loading direction (for uni-axial loading) and the orthogonal direction (for bi-axial loading). The distance vectors for a given periodic boundary pair in its deformed and undeformed location are given by $\bb_i$ and $\bB_i$. \revision{We set $\omega_\epsilon=10^{6}$ for all our experiments. The constraint violation is practically zero ($5.432\times 10^{-5}\%$ on average) and we observe no failure cases for convergence.}
Finally, we use the logarithmic barrier potential $E_{\mathrm{contact}}$ with an adaptive weighting strategy proposed by Li \etal~\shortcite{li2020incremental} to handle self-contacts. Note that contacts need to be resolved both within and across the simulation unit. For this purpose, we compute $E_{\mathrm{contact}}$ with a $2\times2$ tiling of the simulation unit, while retaining the DoF of a single unit. The corresponding derivatives are trivial to compute for this linear map. The total energy is minimized with Newton's method augmented with a back-tracking line search strategy that ensures inversion- and intersection-free steps~\cite{li2020incremental} and adaptive diagonal regularization on the Hessian.

\subsection{Macroscale Homogenization}
\label{sec:method_homo}

\paragraph{Homogenization}
We describe the macromechanical behavior of our metamaterials through the second Piola-Kirchhoff stress tensor $\bS$ as a function of the Green-Lagrange strain tensor $\bE$. These two quantities are work-conjugate \cite{bonet1997nonlinear}, \ie, they satisfy the differential relation 
\begin{equation}
    \bS(\bE) = \frac{\partial \Psi(\bE)}{\partial \bE} \ ,
\end{equation}
where $\Psi$ is the strain energy density function. While we use the Green-Lagrange strain and the second Piola-Kirchhoff stress, our formulation readily extends to other work-conjugate pairs such as deformation gradient and first Piola-Kirchhoff stress.
\begin{figure}[h!]
    \centering
\includegraphics[width=\linewidth]{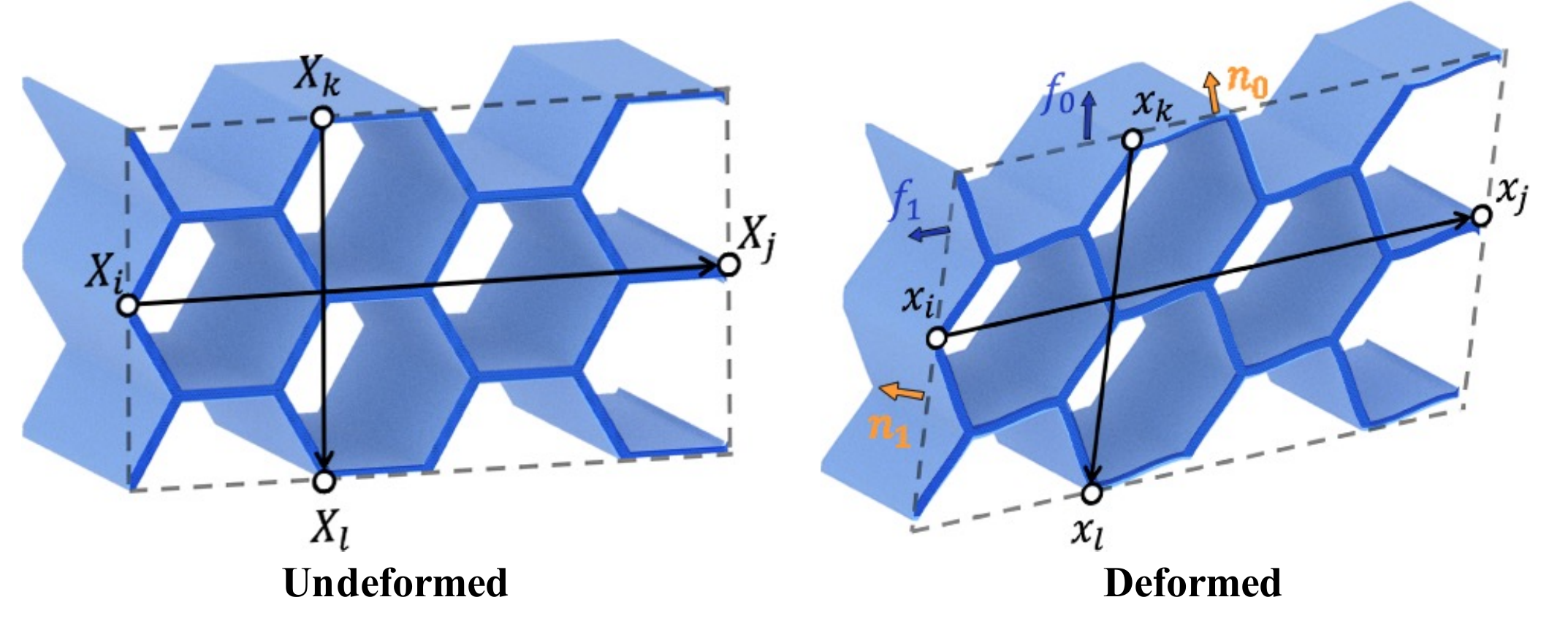}
    \caption{Homogenization. We compute the deformation gradient from corresponding vertex pairs on opposite boundaries of the simulation patch in its deformed (right) and rest (left) state. The averaged Cauchy stress is computed from the internal forces on the boundaries and their corresponding normals.}
    \label{fig:periodic_unit}
    
\end{figure}
We compute the deformation gradient $\bF$ and the Cauchy stress $\pmb{\sigma}$ from the boundary of the simulation patch as
\begin{equation}
\begin{aligned}
    \bF &= \begin{bmatrix} \bx_j-\bx_i \quad \bx_l-\bx_k \end{bmatrix}\begin{bmatrix} \bX_j-\bX_i \quad \bX_l-\bX_k \end{bmatrix}^{-1}, \\
    \pmb{\sigma} &= \begin{bmatrix} \bff_0 \quad \bff_1\end{bmatrix}\begin{bmatrix} \mathbf{n}_0 \quad \mathbf{n}_1\end{bmatrix}^{-1} \ ,
\end{aligned}
\end{equation}
where $\bx$ and $\bX$ denote the deformed and undeformed locations of corresponding vertex pairs $i,j$ and $k,l$ on opposite boundaries of the simulation patch. Furthermore, $\bff_0, \bff_1$ and $\mathbf{n}_0, \mathbf{n}_1$ are the forces and normal directions on adjacent boundaries. See also Fig.~\ref{fig:periodic_unit}.
Having computed these macroscale quantities, the corresponding Green-Lagrange strain and second Piola-Kirchhoff stress follow as
\begin{equation}
    \bE = \frac{1}{2}(\bF^{\tran}\bF - \mathbf{I}) \ , \quad  \bS = J\bF^{-1}\pmb{\sigma}^{{\tran}}\bF^{-{\tran}} \ ,
\end{equation}
where $J=\det\bF$. Finally, the energy density of the unit patch is computed as
\begin{equation}
    \Psi = \frac{E_{\mathrm{total}}}{V} \ ,
\end{equation}
where $V$ is the volume of the simulation patch.
\par

\paragraph{Data Generation}
To sample a given material's stress-strain response, we impose macroscopic strains corresponding to uni- and bi-axial loading with different directions and magnitudes. Although uni-axial loading could be considered a special case of bi-axial loading, it is arguably the scenario that occurs most frequently in applications. It is also worth noting that uni-axial loading generally leads to bi-axial strain: the \textit{lateral} strain orthogonal to the \textit{axial} load is the one that minimizes the elastic energy---and, depending on the material's Poisson ratio, the energetically-optimal lateral strain is generally nonzero. 
\par
Instead of uniformly sampling bi-axial load cases, we, therefore, create a dedicated set corresponding to uni-axial loading such as to properly capture this characteristic material response. 
We start by uniformly sampling directions ranging from $0$ to $\pi$. For uni-axial loading, we create a 1D grid of samples along the given direction, ranging from $30\%$ compression to $50\%$ tension. For bi-axial loading, we generate a 2D grid of equidistant samples ranging from $10\%$ compression to $20\%$ tension along either axis. 
For a given set of tiling parameters, we thus run 2,250 simulations in total. 
\par
When sampling the tiling parameters, we first define a region of interest in parameter space such that no self-intersection occurs in the unit cell
and perform uniform sampling within this range. Depending on the number of structure parameters, we arrive at a total of 900,000 to 4,500,000 simulations per metamaterial family.
\subsection{Neural Metamaterials}
\label{sec:method_ncm}

With the simulation data in hand, our goal is to find a smooth representation of the macromechanical behavior across entire metamaterial families.
Interpolating stress-strain data with radial basis function is an option that has been explored in the past \cite{bickel2009capture}. While we can expect reasonable behavior close to samples, stresses will generally not be conservative away from the data and it is unclear how to estimate, or even bound, integrability error a-priori.
To avoid these problems by construction, we turn to machine learning and use deep neural networks to learn a differentiable energy density function whose derivatives best approximate the stress-strain data obtained from simulation. Training this network on simulation data from many different structures means that the resulting energy density function will describe the constitutive behavior across entire metamaterial families. 
This neural representation---which we term Neural Metamaterial Networks (NMN)---has two important advantages. First, stresses are obtained from the network through standard back-propagation. Since these stresses derive from a potential, they are guaranteed to be conservative everywhere---even away from data points. 
Second, the network is agnostic to parametric discontinuities that would arise due to mesh connectivity changes in native-scale simulations. 
Consequently, the derivatives of the network with respect to structure parameters are smooth and, as discussed in Sec. \ref{sec:method_inv_design}, can thus be used for inverse material design. 
\par
\begin{wrapfigure}{r}{0.25\textwidth} 
    \centering
    \includegraphics[width=\linewidth]{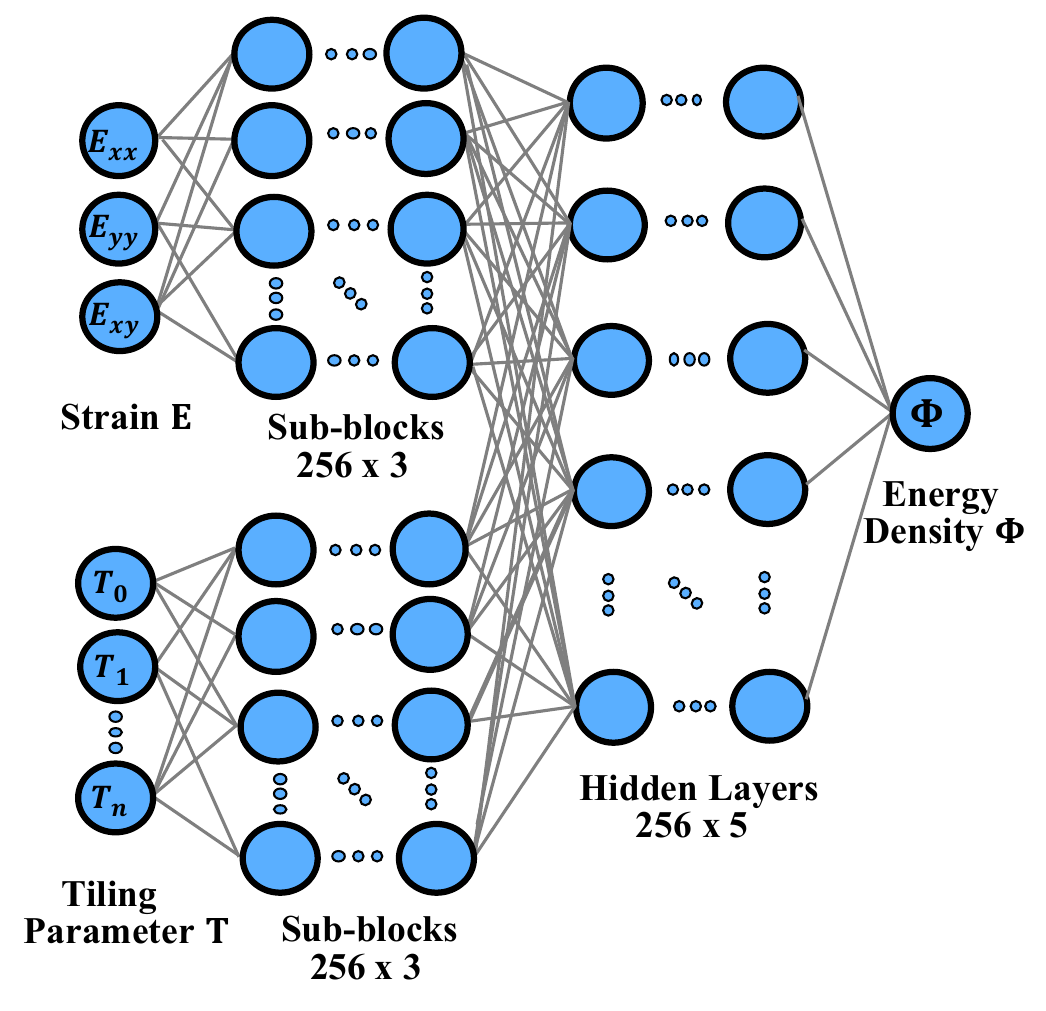}
    \caption{Network Architecture. We use a two-branch MLP that maps strain and tiling parameters to energy density. }
    \label{fig:nn}
\end{wrapfigure}

As illustrated in  Fig.~\ref{fig:nn}, our Neural Metamaterial Network uses a two-branch Multi-Layer Perceptron (MLP) architecture that receives 
tiling parameters $\bT$ and strain $\bE$ as separate inputs and returns the corresponding energy density $\Phi$ as output. 
Both inputs are fed into sub-blocks of 3 layers and then concatenated together to pass through the remaining layers. We use 256 neurons and Swish activation functions~\cite{ramachandran2017searching}, \ie, $x\rightarrow x\cdot\mathrm{sigmoid}(x)$, for all layers except the final one,  which uses a \textit{softplus} activation function, $x\rightarrow\log(1+\exp(x))$, to ensure positive energy. We refer to Sec.~\ref{sec:activation} for an ablation study on the influence of different activation functions.
\par
For training, we use both energy densities $\Psi_i=\Psi(\bE_i)$ and stress-strain pairs $(\bS_i,\bE_i)$ from the simulation data and minimize the loss function 
\begin{equation}
    \label{eq:ncm}
    \begin{aligned}
        \min_{\pmb{\theta}}\:\quad &L_\mathrm{NN}(\pmb{\theta}, \bE, \bT, \pmb{\Psi}) \\
        &= \frac{1}{N_{\mathrm{sp}}}\sum_{i=0}^{N_{\mathrm{sp}}}\frac{1}{\Psi_i} \left(\Phi(\mathbf{T_i}, \bE_i, \pmb{\theta})) - \Psi_i \right)^2\\ &+ \frac{1}{N_{\mathrm{sp}}}\sum_{i=0}^{N_{\mathrm{sp}}} \left|\left|\frac{1}{||\bS_i||_2} \left( \frac{\partial \Phi(\bT_i, \bE_i, \pmb{\theta})}{\partial \bE_i} - \bS_i \right)\right|\right|_2^2,
    \end{aligned}
\end{equation}
with respect to the network parameters $\pmb{\theta}$. The first and second terms in this expression penalize the differences in energy and gradient across all $N_{sp}$ samples. 
We further normalize each sample's contribution with respect to its target values, thus ensuring proper relative weighting for different strain magnitudes.

\subsection{Optimization-based Inverse Design}
\label{sec:method_inv_design}
Neural Metamaterial Networks offer a smooth representation of the macromechanical stress-strain behavior of a given metamaterial family. 
This description directly supports interactive explorations of the parameter space with instant feedback on the nonlinear mechanics of any material inside the corresponding family.
What is more, derivatives of energy density and stress-strain behavior with respect to structure parameters can be obtained in closed form through simple back-propagation. As we describe below, this gradient information opens the door to optimization-based material design algorithms that automatically compute tiling parameters such as to best approximate macromechanical performance goals. 
\paragraph{Uniaxial Stress-Strain Profile.}
A basic inverse design task is to ask for a material whose strain-stress curve best interpolates a set of target pairs $(E_{\alpha,i},S_{\alpha,i})$ under uniaxial loading in a given direction $\alpha$. We formulate this task as a bi-level optimization problem 
\begin{subequations}
\label{eqn:strain_stress}
\begin{align}
\begin{split}
\hspace{-50pt} \min_{\bT}\: O(\bT, \bE(\bT)) = \sum_{i} \left( \left(\bd_\alpha^{\tran}\frac{\partial \Phi(\bT, \bE_{\alpha, i}(\bT))}{\partial \bE_{\alpha, i}}\bd_\alpha\right) - S_{\alpha, i} \right)^2
\end{split}\\
\begin{split}
\hspace{-35pt}\text{s.t.} \quad \bT_{\mathrm{min}} \leq \bT \leq \bT_{\mathrm{max}},
\end{split}\\
\begin{split}
\label{eqn:constraint0}
\quad \quad \quad \bE_{\alpha, i} &= \underset{\bE^*_{\alpha}}{\mathrm{argmin}} \: \Phi(\bE^*_\alpha, \bT), \quad \forall i\\
&\text{s.t.} \quad \bd_\alpha^{\tran}\bE^*_{\alpha}\bd_\alpha = E_{\alpha, i}
\end{split}
\end{align}
\end{subequations}
where $\bT_{\mathrm{min}}, \bT_{\mathrm{max}}$ are the parameter bounds used to generate the training data. The outer objective penalizes deviations from the target stress $S_{\alpha, i}$, whereas the inner optimization ensures that the associated strains $\bE_{\alpha,i}$ have the prescribed magnitude $E_{\alpha, i}$ in the direction of loading while minimizing the elastic energy orthogonal to it. 

\paragraph{Directional Stiffness Profile.} 
Instead of prescribing a particular strain-stress behavior in a given direction, we can also specify directional stiffness profiles. Working with these directional stiffness profiles also enables control over isotropic, orthotropic, or completely anisotropic behavior. 
We cast directional stiffness design as another bi-level optimization problem 
\begin{subequations}
\begin{align}
\begin{split}
\hspace{-50pt} \min_{\bT}\: O(\bT, \bE(\bT)) \nonumber
\end{split}\\
\begin{split}
\label{eqn:stiffness_opt}
\hspace{-50pt} = \sum_{\alpha} \left( \left((\bd_\alpha \bd_\alpha^{\tran}) : \mathbb{S}(\bT, \bE_\alpha(\bT))_\alpha : (\bd_\alpha \bd_\alpha^{\tran})\right)^{-1} - k_{\alpha} \right)^2,
\end{split}\\
\begin{split}
\hspace{-32pt}\text{s.t.} \quad \bT_{\mathrm{min}} \leq \bT \leq \bT_{\mathrm{max}},
\end{split}\\
\begin{split}
\label{eqn:constraint1}
\quad \quad \quad \bE_\alpha &= \underset{\bE^*_\alpha}{\mathrm{argmin}} \: \Phi(\bE^*_\alpha, \bT), \quad \forall \alpha\\
&\text{s.t.} \quad \bd_\alpha^{\tran}\bE^*_\alpha \bd_\alpha = E_\alpha
\end{split}
\end{align}
\end{subequations}
in which the outer objective aims to match the directional stiffness targets $k_\alpha$ and $\mathbb{S}_\alpha$ is the compliance tensor for direction $\alpha$. The inner objective again ensures that strain corresponds to uni-axial loading.  
It should be noted that the compliance tensor, which is defined through
\begin{equation}
    \mathbb{S}_\alpha:\mathbb{C}_\alpha = \mathbb{I} \quad \ , \quad  \mathbb{C}_\alpha = \frac{\partial \Phi^2(\bT, \bE_\alpha)}{\partial \bE_\alpha^2} \ ,
\end{equation}
involves second derivatives of the network. In practice, we compute $\mathbb{S}_\alpha$ by inverting $\mathbb{C}_\alpha$ in matrix form.
\paragraph{Poisson Ratio Profile.}
%
Optimizing for stiffness curves along given directions, or profiles spanning all directions, corresponds to designing for generalized Young's moduli. We can likewise define design objectives for generalized Poisson's ratios that control the contraction (or expansion) rate in the directions orthogonal to uni-axial loading.
To this end, we pose another optimization problem,
\begin{equation}
    \begin{aligned}
        \hspace{-20pt} & \min_{\bT}\: O(\bT, \bE(\bT)) \\
        \hspace{-20pt} &= \sum_{\alpha} \left( -\frac{(\bd_\alpha \bd_\alpha^{\tran}) : \mathbb{S}(\bT, \bE_\alpha(\bT))_\alpha : (\bn_\alpha \bn_\alpha^{\tran})}{(\bd_\alpha \bd_\alpha^{\tran}) : \mathbb{S}(\bT, \bE_\alpha(\bT))_\alpha : (\bd_\alpha \bd_\alpha^{\tran})} - \nu_{\alpha} \right)^2,
    \end{aligned}
\end{equation}
where $\nu_\alpha$ are target values for Poisson's ratio and $\bn_\alpha$ is the direction orthogonal to $\bd_\alpha$. 
It should be noted that this expression is subject to the constraints defined in Eq.~\ref{eqn:stiffness_opt}, which we omitted here for brevity. 

\paragraph{Optimization.}
The constrained optimization problems introduced above can be expressed in general form as
\begin{equation}
\begin{aligned}
    & \min_{\bT}\: O(\bT, \bq(\bT))\\
     & \text{s.t.} \quad \bg(\bT, \bq(\bT)) = \bZero,
\end{aligned}
\end{equation}
where $\bg$ is a vector-valued constraint function, and $\bq = [\bE, \pmb{\lambda}]^{\tran}$ is the concatenation of strain variables and Lagrange multipliers associated with the directional strain magnitude constraints.
We solve these optimization problems using a combination of sensitivity analysis for the outer problem and sequential quadratic programming (SQP) for the inner problem.
To this end, we express the total derivative of the objective function as
\begin{equation}    
 \label{eq:totalDerivativeObjective}
 \frac{\dd O}{\dd \bT} = \frac{\partial O}{\partial \bT} + \frac{\partial O}{\partial \bq}\frac{\dd \bq}{\dd \bT} \ ,
\end{equation}
where we have used the fact that equilibrium strain and Lagrange multipliers are completely determined by the choice of tiling parameters and we can therefore write $\bq=\bq(\bT)$. While the partial derivatives in (\ref{eq:totalDerivativeObjective}) are straightforward to compute using auto-differentiation, the remaining total derivative requires more attention. The relation between equilibrium $\bq$ and $\bT$ is implicitly given through the inner optimization problem. To obtain an expression for this term, we start by writing out the Lagrangian of the inner optimization problem as
\begin{equation}
    \Lag(\bq) = \sum_\alpha \Phi(\bE_\alpha^*) - \lambda_\alpha (\bd_\alpha^{\tran}\bE_\alpha^*\bd_\alpha - E_\alpha) \ .
\end{equation}
The first-order optimality conditions require that the gradient of the Lagrangian vanish at the optimum, \ie,  
\begin{equation}
    \bg := \frac{\dd \Lag}{\dd \bq} = \bZero \ .
\end{equation}
Differentiating both sides \wrt the tiling parameters $\bT$, we obtain the closed-form expression 
\begin{equation}
    \frac{\dd \bg}{\dd \bT} = \frac{\partial \bg}{\partial \bT} + \frac{\dd \bg}{\dd \bq}\frac{\dd \bq}{\dd \bT} = 0 \ ,
\end{equation}

\begin{equation}
    \frac{\dd \bq}{\dd \bT} = -\frac{\dd \bg}{\dd \bq}^{-1}\frac{\partial \bg}{\partial \bT} \ .
\end{equation}

\par
With the total derivative of the design objective in hand, we can leverage quasi-Newton methods for efficient, gradient-based minimization. We choose the L-BFGS-B method with a backtracking line search, which proved a robust and efficient choice for all our examples. Whenever we evaluate the objective function or its gradient, we must first solve the inner optimization problem. We use sequential quadratic programming (SQP) for this purpose, with analytical gradient and Hessian of the objective function obtained via network auto differentiation. For increased robustness, we additionally enforce the Hessian to be positive definite. Since the size of this matrix is relatively small, we perform eigen decomposition and shift eigenvalues such that the smallest one is equal to $10^{-7}$.

\section{Results}
We analyze our Neural Metamaterial Networks on a set of examples that highlight the advantages of this smooth neural representation for inverse material design.
We first report the training statistics for multiple metamaterial families (Sec.~\ref{sec:ncm}). 
Leveraging these neural representations, we then perform inverse design for strain-stress profile, stiffness, and Poisson ratio optimization (Sec.~\ref{sec:inverse}).
Finally, we conduct ablation studies on the benefits of the smooth neural representation and different choices of network activation functions (Sec.~\ref{sec:ablation}).
\subsection{Training}
\label{sec:ncm}
We use Adam as our optimizer with a learning rate of $10^{-4}$ for all of our experiments. Training is performed using a \textit{GeForce RTX 3080} with a batch size of \new{40,000}, and we train for 8,000 epochs. 
We set aside 5\% of the simulation data for testing, whereas the remainder is used for training the network. 
Test errors are summarized in Table~\ref{tab:stats}, with energy and gradient error measuring the relative difference in energy densities and the second Piola-Kirchhoff stresses.
With a maximum test error of $1.63\%$ across all families, we conclude that our approach is well-suited to represent the macromechanical behavior of metamaterial spaces.
Depending on the number of structure parameters, the training time can vary between 2 to 4 days. 
\par
In addition to validations on the test set, we further analyze the accuracy of the network predictions using uni-axial loading tests. We create a separate test set by randomly sampling 50 combinations of different structure parameters and loading directions per family. We use strains in the range of $30\%$ compression and $50\%$ tension at 25 uniformly distributed locations, \ie 1,250 testing points per family. We evaluate the relative error between the directional stress computed from optimizing over the network and native-scale simulation. We obtain average and maximum errors of $0.475\%$ and $3.57\%$ across all families, which we consider sufficiently accurate for inverse design. We refer to the supplementary material for the directional strain-stress curves for all test cases. 

\begin{table}[h!]
\caption{Error evaluated on the test set for different tiling families. As can be seen from the two rightmost columns, we consistently achieve low error for both energy and gradient loss.}  
\label{tab:stats}
\begin{tabular}{ |c|c|c|c| } 
\hline
Tiling Family & \# Params  & Rel. Gradient Err. & Rel. Energy Err. \\
\hline
IH29 & 1  & 1.38\% & 0.709\% \\ 
\hline
IH21 & 2  & 1.61\% & 0.883\% \\ 
\hline
IH50 & 2  & 1.14\% & 0.921\% \\ 
\hline
IH67 & 2  & 1.63\% & 0.739\% \\ 
\hline
IH28 & 2  & 1.05\% & 0.472\% \\ 
\hline
IH22 & 3  & 1.57\% & 1.13\% \\ 
\hline
IH01 & 4  & 1.20\% & 0.579\% \\ 
\hline
\end{tabular}
\end{table}
\begin{figure*}[ht]
    \centering
    \includegraphics[width=\linewidth]{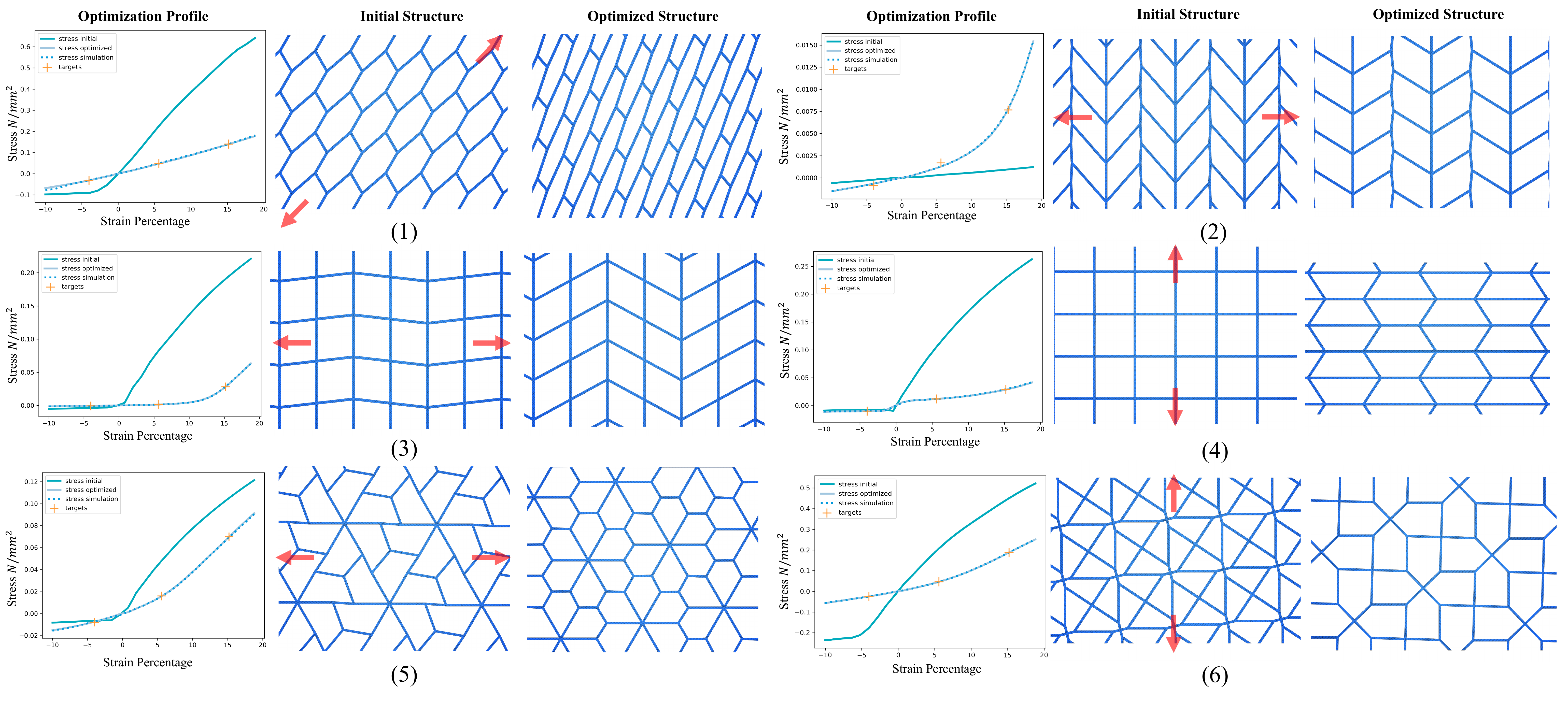}
    \caption{Optimization of Strain-Stress Curves. Here we perform inverse design on different tiling families to obtain structures whose strain-stress curve passes through prescribed targets. 
    \new{Orange crosses mark target points, and cyan curves indicate the initial profile. Solid sky-blue curves show the optimized structure, while the dotted blue curves plot the reference from native-scale simulation.
    We showcase examples of mapping between highly nonlinear and quasi-linear profiles (first row), significantly decreasing stiffness for tension (second row), and modifying the mechanical responses under both tensile and compressive loading (third row).}
    The uni-axial loading directions are indicated with red arrows.
    }
    \label{fig:strain_stress_opt}
\end{figure*}
\begin{figure*}[ht]
    \centering
    \includegraphics[width=\linewidth]{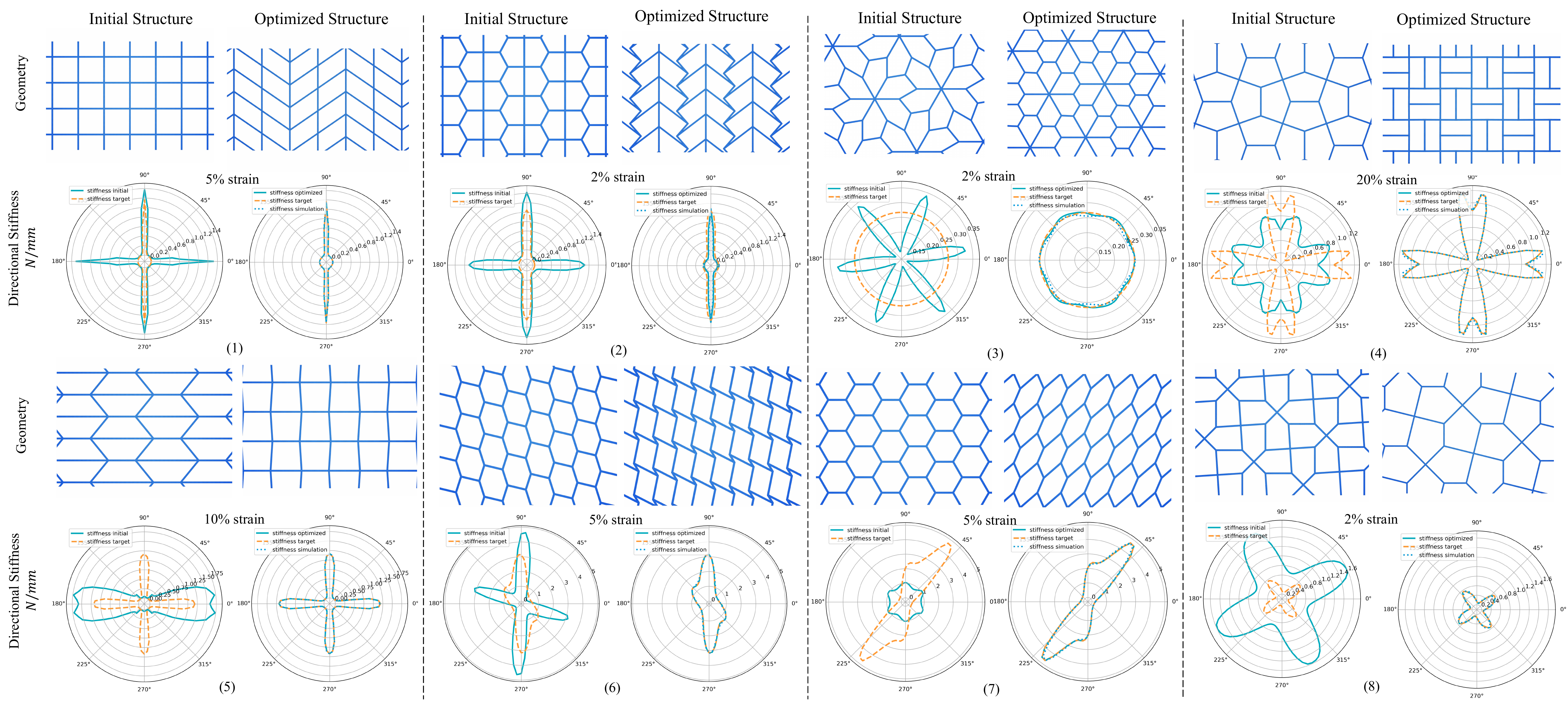}
    \caption{Directional Stiffness Optimization.
    We optimize for structure parameters such as to transform initial stiffness profiles (cyan) into given target profiles (orange).
    \new{As can be seen that, while showing good agreement with their simulation counterparts (dotted blue curves), our approach performs robustly for various types of design tasks, \eg significantly changing the stiffness for an orthotropic material (\textit{1, 2}), transforming an anisotropic material into nearly isotropic (\textit{3}) and orthotropic (\textit{4}) ones, and mapping between different anisotropic targets (\textit{5--8}). }
    }
    \label{fig:stiffness_opt}
\end{figure*}
\begin{figure*}[ht]
    \centering
    \includegraphics[width=\linewidth]{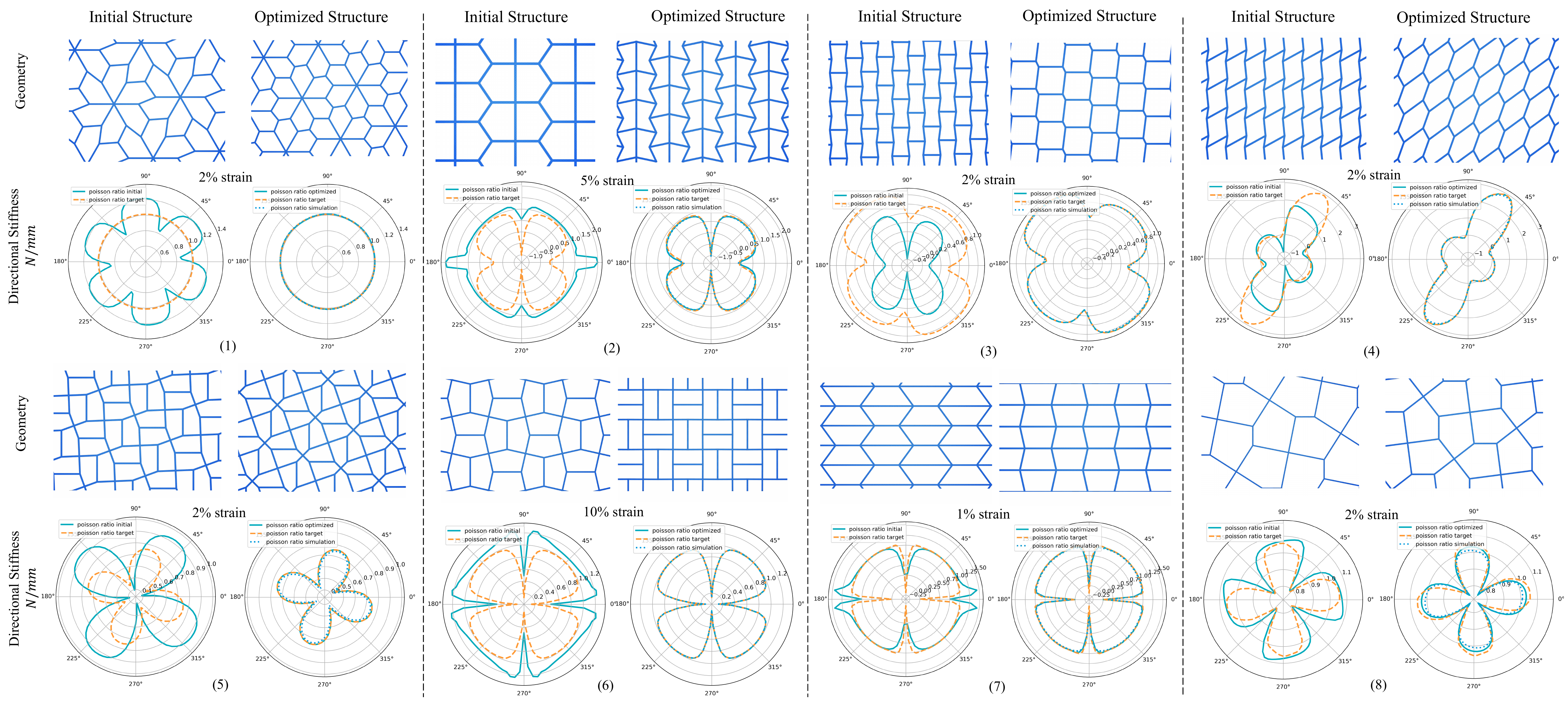}
    \caption{Poisson Ratio Optimization.
       We optimize for structure parameters such as to transform initial Poisson ratio profiles (cyan) into given target profiles (orange). \new{It can be seen that optimized structures closely match the prescribed target and show good agreement with the simulation reference (dotted curves).}
    }
    \label{fig:poisson_ratio_opt}
\end{figure*}
\subsection{Inverse Design}
\label{sec:inverse}
We analyze the ability of our network to perform inverse design tasks. To this end, we select three representative criteria that are intuitive as user inputs: uni-axial stress, directional stiffness, and directional Poisson ratio, as defined in Sec~\ref{sec:method_inv_design}.
All tasks can be formulated as bi-level constrained optimization problems, in which we leverage the analytical derivatives from auto differentiation over the network.
In the first example, we optimize for the strain-stress behavior in a given direction for uni-axial loading, whereas in the second and third tasks, we optimize for the directional stiffness and Poisson ratio profile for a given strain magnitude. 
\new{For all examples discussed in this section, we compute the ground-truth values for the optimized structures using native-scale simulation (shown as dotted curves). These comparisons indicate that our neural representation offers very good agreement with the simulation reference.}
\paragraph{Uni-axial Stress.} 
In the set of examples shown in Fig.~\ref{fig:strain_stress_opt}, we apply uni-axial loading along a given direction and seek optimal tiling parameters such that the stresses for different strain magnitudes match prescribed target values. 
\new{In the examples shown in the first row, we set directional stress targets such as to map between nonlinear and quasi-linear profiles. }
For the two examples on the second row of Fig.~\ref{fig:strain_stress_opt}, we set targets that significantly decrease stress in the given directions. The optimized structures change stress by creating geometries that modify the deformation range compared to the initial structure.
For the design target in the bottom row, we decrease stiffness both under compression and tension.
As can be seen in all these examples, tiling geometries change significantly to minimize the design objectives. \new{The strain used for these examples is within $10\%$ compression and $20\%$ tension. }
%
%
\paragraph{Directional Stiffness.}
In a second series of experiments, whose results are shown in Fig.~\ref{fig:stiffness_opt}, we investigate the ability of our method to optimize for directional stiffness profiles. \new{For these tests, we impose tensile strains between $2\%$ and $20\%$}
In the first and second examples, we start with orthotropic materials and ask for a significant decrease in stiffness in one of the principal directions. As can be seen from the resulting geometry, the horizontal beams fold to allow for a larger range of low-energy deformations, thus reducing stiffness in the corresponding direction.
For the third experiment, we set an orthotropic profile as a target and observe vertical beams straightening such as to increase stiffness.
In the remaining examples, we transform an initially anisotropic material into an isotropic and orthotropic one, and map between different anisotropic targets. 

\begin{figure}[h!]
    \centering
    \includegraphics[width=\linewidth]{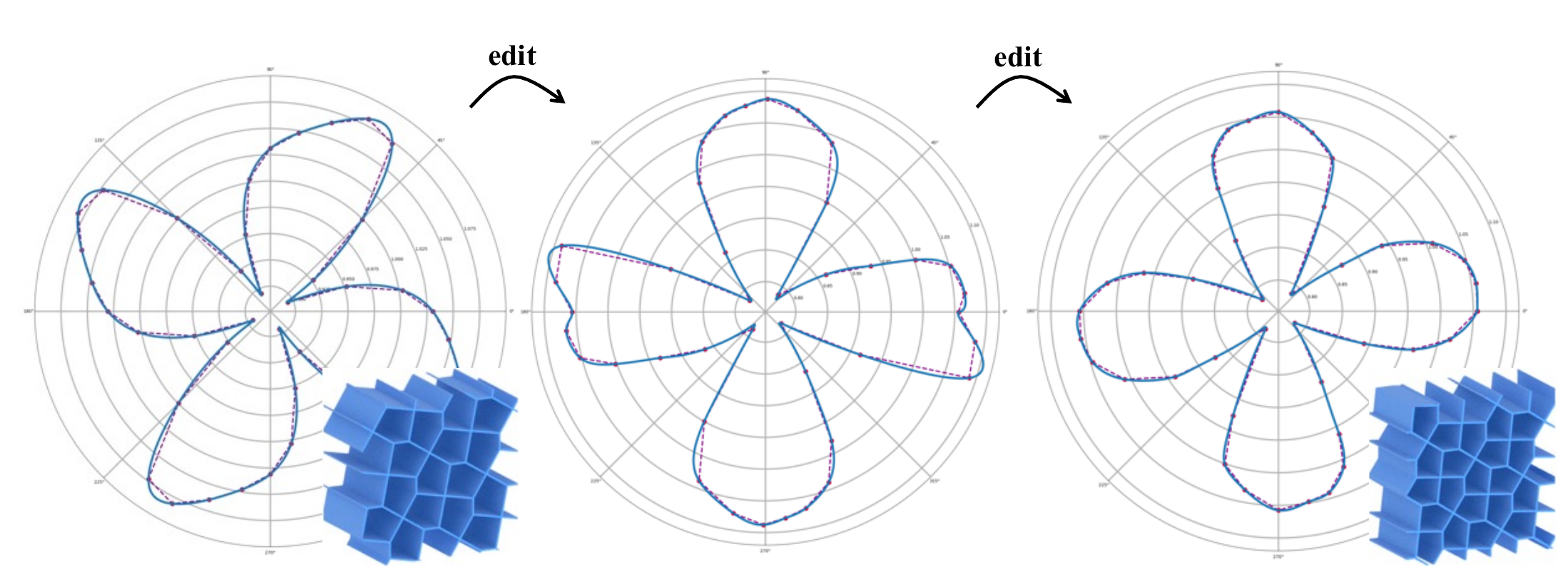}
    \caption{Interactive Target Profile Modifier. The user can edit target profiles by modifying the control points of an underlying spline representation. The initial and optimized structures are shown in the bottom right corners.}
    \label{fig:target_gen}
\end{figure}

%
\paragraph{Directional Poisson Ratio.}
Complementing the previous examples for stiffness design in the direction of loading, we now turn to lateral coupling in terms of Poisson's ratio.
\new{
As can be seen from Fig.~\ref{fig:poisson_ratio_opt}, our method successfully turns a high-frequency profile into an isotropic one (first example), transforms Poisson ratios in a given direction from positive to negative (second example), and vice versa (third example). The remaining examples further showcase our approach on a range of anisotropic targets. 
\new{We impose tensile strains between $1\%$ and $10\%$ for these tests.}
}
%
\paragraph{Target Specification.}
In order to specify target profiles in an intuitive and convenient way, we built a simple interface that allows users to change the control points of an underlying spline representation.
Fig.~\ref{fig:target_gen} demonstrates the process of creating the Poisson ratio target for the last example (\textit{8}) shown in Fig.~\ref{fig:poisson_ratio_opt}. 
\paragraph{Timings.}
Our network architecture allows for real-time inference. Specifically, the average time for evaluating the energy and its first and second derivatives are $3.26\times 10^{-4}s$, $2.14 \times 10^{-4}s$ and $2.77\times 10^{-4}s$, respectively. The timings for the inverse design examples can vary depending on the objective and initial parameters provided by the user. For brevity, we only report timings for the first examples shown in Fig.~\ref{fig:strain_stress_opt}, ~\ref{fig:stiffness_opt}, and ~\ref{fig:poisson_ratio_opt}, which are \new{$3.93s$, $49.3s$ and $36.1s$}, respectively. These timings were obtained on a workstation with a \textit{GeForce RTX 3080} GPU and an \textit{AMD Ryzen 9 5950X} CPU. The simulations used to generate training data are performed in parallel on a cluster. Although the exact timing depends on the available computation resources, the time for generating the training data used in this paper is roughly the same time required for training the corresponding networks.
\paragraph{Implementation Details.}
Our customized simulator is implemented in C++ using Eigen~\cite{eigenweb} for linear algebra operations, Intel TBB for parallelization, and CHOLMOD~\cite{chen2008algorithm} for solving linear systems. To create microstructure geometry from parameters, we use the Clipper2 library\footnote{https://github.com/AngusJohnson/Clipper2} to assign thickness to the wired edges, and Gmsh~\cite{geuzaine2009gmsh} for boolean operations as well as generating periodic meshes. These two procedures together take on average $1.2s$. 

%
\subsection{Ablation Study}
\label{sec:ablation}
\begin{figure}[h!]
    \centering
    \includegraphics[width=\linewidth]{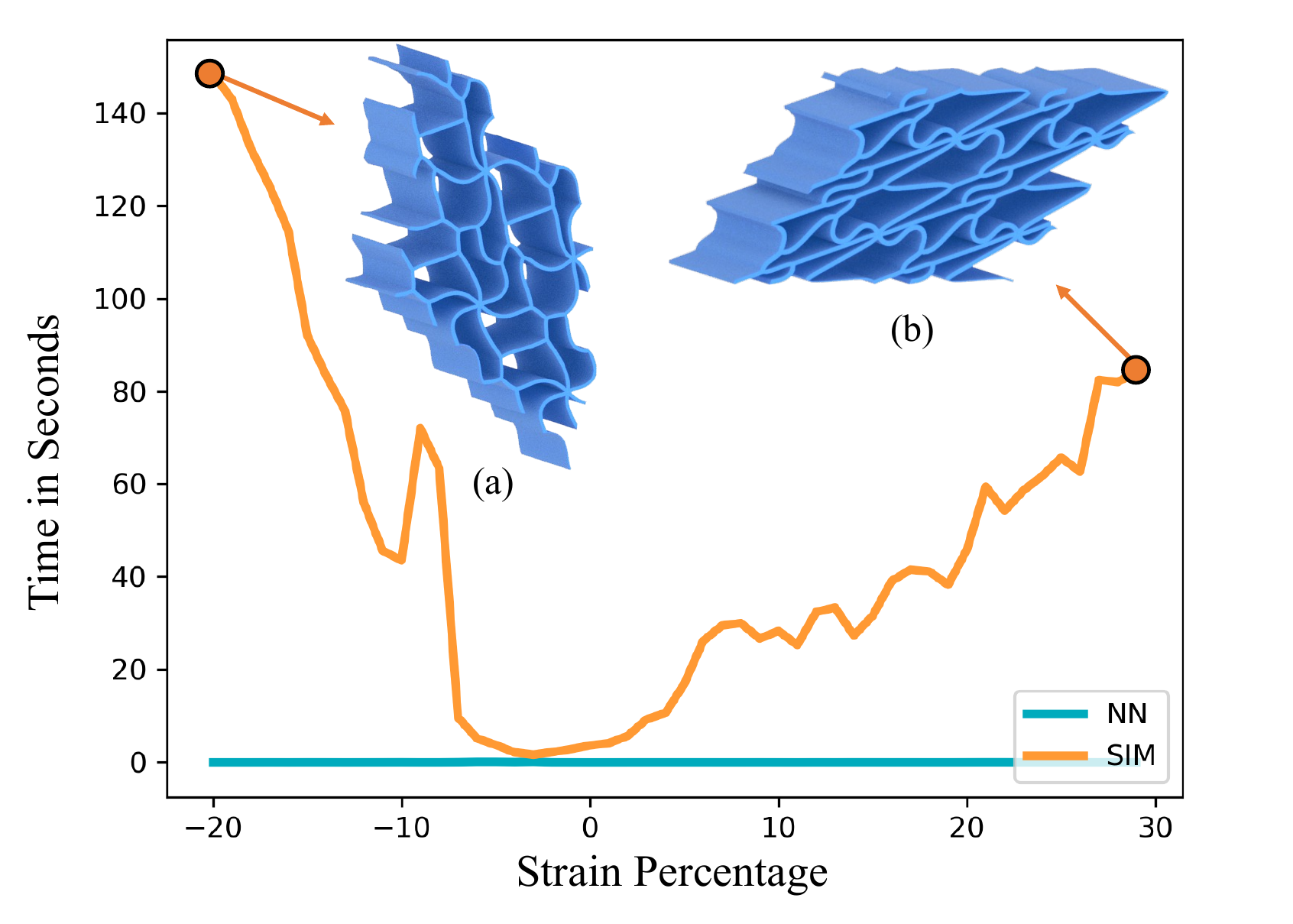}
    \caption{
    \new{
    Timing comparison between network prediction and native-scale simulation. We compare the time required for finding equilibrium states under uni-axial loading using our neural representation (cyan curve) and native-scale simulation (orange curve). Optimizing over our neural representation yields consistently lower computation time ($0.077s$ on average) for finding equilibrium states. This comes with no surprise as the native-scale simulation has to cope with compression (a) and contacts (b) under large deformation, The performance of native-scale simulation varies strongly with loading conditions and can take up to $148s$. 
    }
    }
    \label{fig:timing}
\end{figure}

\paragraph{Comparsion to Native-scale simulation.}
All our inverse design examples involve an inner loop that ensures uni-axial loading. Unlike simple network evaluation, this process requires solving an optimization problem as in Eqs. (\ref{eqn:constraint0}) and  (\ref{eqn:constraint1}). Nonetheless, our neural representation offers significant computational advantages compared to native-scale simulation. Since our approach targets large deformations, it is likely that the native-scale simulation will encounter numerically challenging cases, \eg when beams are under compression or contact. These cases, on the other hand, do not pose any additional challenges for the learned model. We compare the computation time for finding the equilibrium states under uni-axial loading using network prediction and native-scale simulation. As can be seen in Fig.~\ref{fig:timing}, the time required for native-scale simulations varies substantially with imposed strain magnitude. The reason for these fluctuations is that the Newton solver can require many iterations for challenging load cases, and each iteration can require several regularization and line search steps.
In contrast, our neural representation exhibits nearly constant computation time, regardless of the loading conditions, thus achieving performance gains of up to $2000\%$. For this example, the underlying mesh has $109,362$ degrees of freedom. 
 
\begin{figure*}[t]
    \centering
    \includegraphics[width=\linewidth]{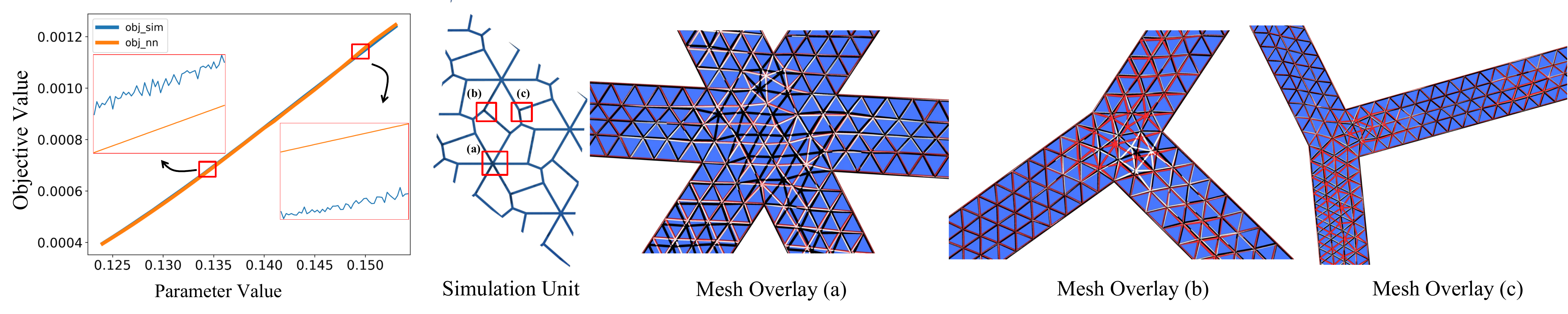}
    \caption{Non-smoothness due to meshing. 
    \textit{From left to right}: objective values obtained by sampling along a given direction in parameter space through both native scale simulations and optimization over the network (\textit{1}). 
    As can be seen from the insets in panel \textit{1}, the objective values from the simulations show clear zig-zag patterns due to the meshing process, whereas our neural metamaterial networks lead to perfectly smooth behavior.
    To visualize the meshing inconsistency, in panels (\textit{3---5}) we overlay the simulation meshes for three consecutive steps for three critical regions of the structure (\textit{2}) with different colors.
    While parameter values only vary by $10^{-6}$, the discretization changes significantly both in nodal positions and topology.
}
    \label{fig:meshing_error}
\end{figure*}

\paragraph{Smooth Neural Representation.} 
\new{
Our method removes the need for native-scale simulation and re-meshing at design time. As one particular advantage, our neural representation provides smoother behavior of macromechanical properties across parameter space compared to simulation-based methods with meshing in the loop.}
To examine this smooth behavior, we sample a given objective function along a given direction in parameter space and compare results obtained through native-scale simulations and our neural representation, respectively. 
This inverse design example is taken from Fig.~\ref{fig:strain_stress_opt}-(\textit{5}). 
As can be seen from Fig.~\ref{fig:meshing_error}, when evaluating the design objective based on native-scale simulations, the plot exhibits high-frequency oscillations around a quasi-linear trend.
To identify the source of this non-smoothness, we overlay the simulation rest state meshes for three consecutive steps in parameter space using a step size of $10^{-6}$. 
The close-up views shown in Fig. \ref{fig:meshing_error} (3--5) reveal that the change in discretization is rather significant compared to the small magnitude of parameter perturbations.
Although our neural representation is trained on mesh-based simulations, it does not require \textit{meshing-in-the-loop} and thus produces smooth behavior across the entire parameter space.
\begin{figure}[h!]
    \centering
    \includegraphics[width=\linewidth]{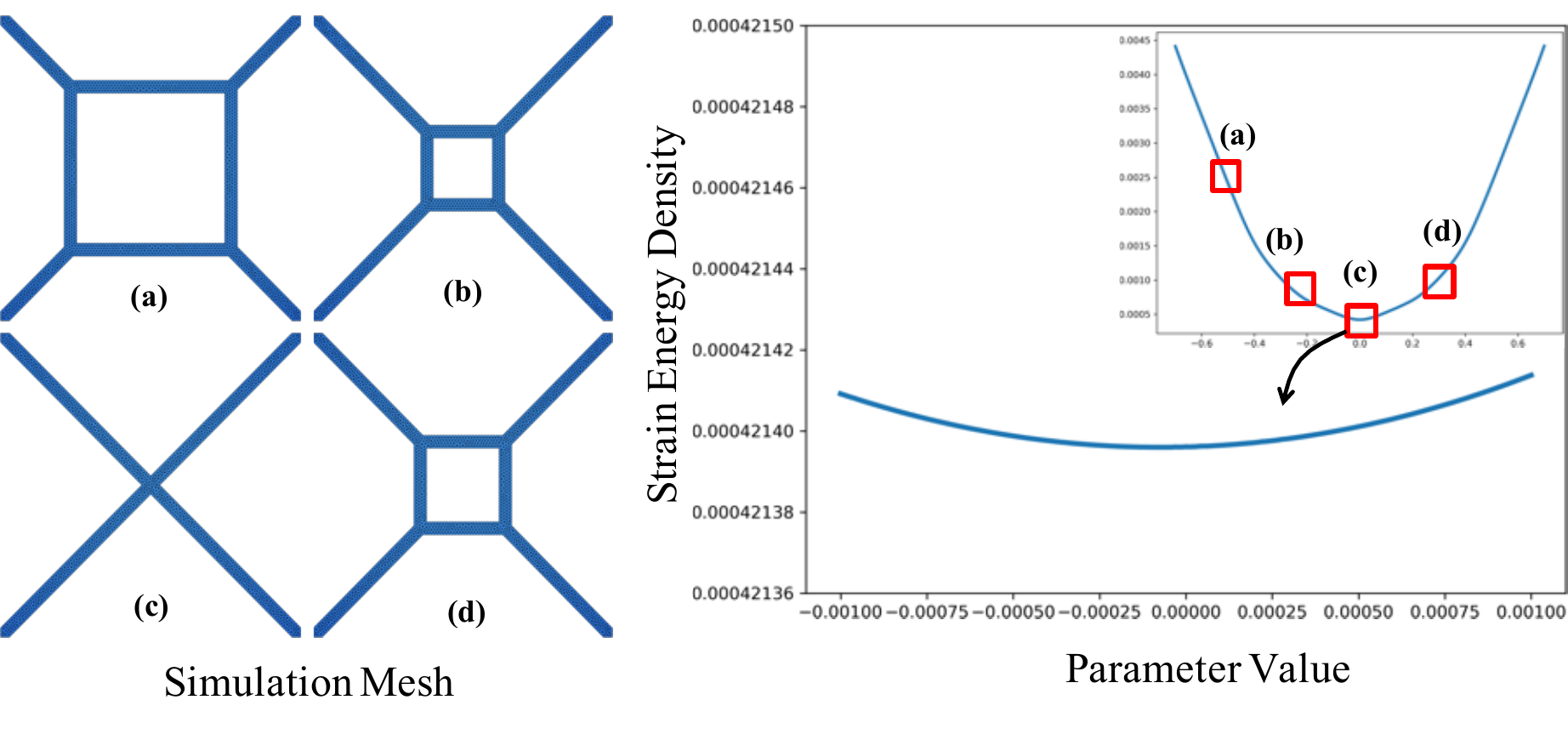}
    \caption{
    Topological Singularity. 
    We consider a metamaterial family whose single parameter $p\in[-1,1]$ controls the size of an inner square as shown in (a---d).
    After training, we probe the macromechanical behavior of this material under uniaxial loading by plotting the energy density from our neural network. We sample across the entire parameter space using a step size of $10^{-5}$.
  While moving through parameter space, the geometry of the structure passes through a singular point (c) where mesh topology must change and native-scale simulation gradients are undefined. As can be seen from the plots on the right our neural representation is smooth even when stepping through the singularity located at $p=0.0$. 
    }
    \label{fig:case_study}
\end{figure}
\par
While adaptive re-meshing and smoothing could alleviate this issue to some extent, they cannot remove non-smoothness for cases where the mesh topology must change to accommodate changes in geometry.
To isolate this problem, we consider a metamaterial family whose parameter space contains a topological singularity. As illustrated in Fig.~\ref{fig:case_study}(a---d), the single parameter of this metamaterial controls the size of an inner square, which first contracts to a single point and then expands again. Shortly before and after the point of full contraction (c), there are two points that correspond to a change in genus. It is evident that there cannot be a mesh of constant connectivity that would be valid for the entire parameter space. As can be seen from the right plot, our neural representation, on the other hand, shows perfectly smooth behavior when passing through this singularity.
Although these discontinuities do not necessarily imply failure for native-scale methods, our neural representation provides a simple and effective way to obtain smooth analytical derivatives that avoid meshing discontinuities from the outset.

\begin{figure}[h!]
    \centering
    \includegraphics[width=\linewidth]{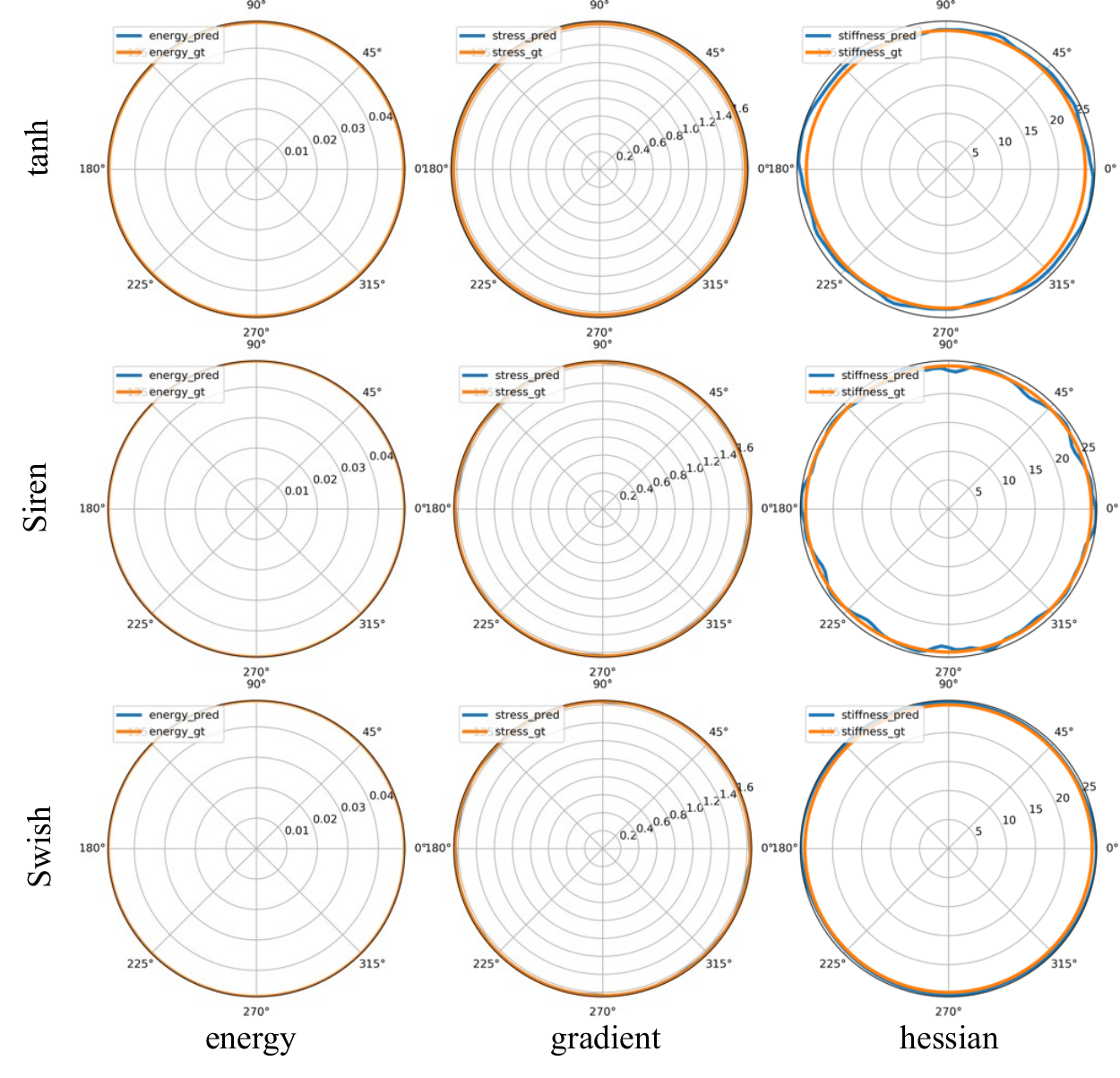}
    \caption{Impact of Activation Functions. We compare three activation functions on the task of fitting an isotropic Neo-Hookean material. While all choices approximate energy and first derivatives accurately, only the Swish activation function produces smooth second derivatives.}
    \label{fig:activation_comparison}
\end{figure}
\paragraph{Network Activation Functions.}
\label{sec:activation}
To study the influence of activation functions, we compare different choices on the simple task of fitting an isotropic Neo-Hookean model~\cite{bonet1997nonlinear}, whose analytical derivatives are readily available.
Specifically, we compare \textit{Swish} (our choice), \textit{tanh}, and \textit{sine} activation functions, all of which provide sufficient smoothness for our application.
While \textit{tanh} is a common choice, sinusoidal activation functions~\cite{sitzmann2020implicit} recently showed promising results in physics-informed learning~\cite{zehnder2021ntopo}.
In Fig.~\ref{fig:activation_comparison}, we show network predictions and ground truth values for the directional energy, gradient, and Hessian for the different activation functions.
Since the underlying constitutive model is isotropic, the material response should be identical for all directions, \ie, all plots should be perfect circles.
While all three choices lead to a very good approximation of energy and gradient values, only the Swish activation function produces smooth second derivatives.
%

\section{Conclusion}
We have presented Neural Metatmaterial Networks---a new approach for representing the nonlinear macromechanics of metamaterial families.
Different from previous methods, we use deep neural networks to represent the macromechanical behavior through a continuous space of structure parameters.
Using isohedral tilings as an example, we demonstrated the potential of our network to learn constitutive relationships for a wide range of materials.
In particular, the smooth analytically-differentiable nature of our network 
enables gradient-based optimization in parameter space, which we demonstrated on various inverse material design tasks. 
We further showed that, by directly learning from macromechanical quantities, our network avoids gradient discontinuities due to meshing that arise when working with native-scale simulations. 
\subsection{Limitations and Future Work}
Our current implementation relies on a comparatively simple sampling strategy. While regular sampling in strain space allows for efficient simulation warm starts, adaptive strategies might be able to improve sample efficiency.
As the number of structure parameters grows, the number of samples required to probe regions of interest quickly explodes. A strategy to mitigate this curse of dimensionality could be to explore reparameterization and dimension-reduction in parameter space such as to focus sampling density where variation in structure is the largest. 
\par
We have used data obtained from native-scale simulations to train our networks. 
While we observed very good agreement between network predictions and simulation baselines, an interesting avenue for future research is to combine synthetic data with real-world measurements. Although energy density cannot be captured from experiments, training only on gradient data is a possibility, in principle. 
\revision{Another direction to explore is to conduct calibration of the native-scale model to experimental data. While we expect our Neo-Hookean material model to provide accurate predictions for actual rubber materials, 3D-printed materials such as thermoplastic polyurethane might require more advanced constitutive models.}
Extending our method to support continuously graded metamaterials is another exciting direction for future work. The parametric space of Voronoi diagrams with star-shaped metrics would be a promising place to start \cite{martinez2019star}. Finally, although we focused on structures extruded from planar patterns in this work, extending our core formulation to 3D would only require a change in the number of strain inputs (from 3 to 6).
We plan to apply our methodology to general 3D metamaterial families, \eg those from Panetta \etal~\shortcite{panetta2015elastic}, in the future.

\begin{acks}
The authors would like to thank Professor Craig S. Kaplan for sharing his isohedral tiling code with the public. We also thank Juan Montes Maestre and Ronan Rinchet for their valuable discussions and suggestions. We further thank Wenzhao Xu for designing the shoe model and for her assistance with the figures. Finally, the authors thank the anonymous reviewers for their valuable feedback. This work was supported by the European Research Council (ERC) under the European Union’s Horizon 2020 research and innovation program (grant agreement No. 866480), and the Swiss National Science Foundation through SNF project grant 200021\_200644.
\end{acks}
\bibliographystyle{ACM-Reference-Format}
\bibliography{reference}
\end{document}